\documentclass{elsarticle}
\usepackage{graphics}
\usepackage{amssymb}
\usepackage{amsmath}
\usepackage{wasysym}
\usepackage{latexsym}
\usepackage{graphicx}
\usepackage{epsfig}
\usepackage{subfigure}
\usepackage{float}
\usepackage{color}
\usepackage{lineno}
\usepackage{verbatim}
\usepackage{enumitem}

\journal{Theory in Biosciences}
\begin{document}


\begin{frontmatter}

\title{Revisiting institutional punishment  in the $N$-person prisoner's dilemma}

 \author[mymainaddress]{Bianca Y. S. Ishikawa}
 \ead{yumibianca@usp.br}

\author[mymainaddress]{Jos\'e F.  Fontanari\corref{mycorrespondingauthor}}
\cortext[mycorrespondingauthor]{Corresponding author}
\ead{fontanari@ifsc.usp.br}
  
\address[mymainaddress]{Instituto de F\'{\i}sica de S\~ao Carlos, Universidade de S\~ao Paulo,  13560-970 S\~ao Carlos, S\~ao Paulo, Brazil}

\begin{abstract}
The conflict between individual and collective interests makes fostering cooperation in human societies a challenging task, requiring drastic measures such as the establishment of sanctioning institutions.  These institutions are costly because they  have to be maintained regardless of the presence or absence of offenders.  Here we revisit some improvements to the standard $N$-person prisoner's dilemma formulation with institutional punishment in a well-mixed population, namely the elimination of overpunishment, the requirement of a minimum number of contributors to establish the sanctioning institution, and the sharing of its maintenance costs once this minimum number is reached.   In addition, we focus on large groups or communities for which sanctioning institutions are ubiquitous.  Using the replicator equation framework for an infinite population, we find that by sufficiently fining players who fail to contribute either to the public good or to the sanctioning institution, a  population of contributors  immune to invasion by these free riders can be established, provided that the  contributors are sufficiently numerous. In a finite population,  we use finite-size scaling to  show  that, for some parameter settings,   demographic noise helps to fixate the strategy that contributes to the public good but not to the sanctioning institution even for infinitely large populations when, somewhat counterintuitively, its proportion in the initial population vanishes with a small power of the population size.
\end{abstract}


\end{frontmatter}

\section{Introduction}\label{sec:intro}

Understanding and promoting cooperation in human societies has been declared one of the greatest challenges for science in the twenty-first century \cite{Kennedy_2005}. The main issue is the conflict between individual and collective interests: either individuals contribute to public goods at a personal cost, or they free ride on the efforts of others \cite{Hardin_1968,Axelrod_1984}. This conflict between different levels of selection, which generally arises when considering social traits, was already noted by Darwin in his discussion of courage and self-sacrifice in humans, since these are traits that are counter-selected within a social group, but increase the group's chances of survival in the case of inter-group fighting \cite{Darwin_1871}. This type of social dilemma has been extensively studied within the  framework  of evolutionary game theory \cite{Maynard_1982}.

The usual evolutionary game theoretical framework  for studying cooperation is the  $N$-person prisoner's dilemma \cite{Fox_1978}. In a play group  of   $N > 1$  individuals, each individual can decide whether to contribute a fixed amount $c> 0$  to the common pool (public goods). This amount is multiplied by a factor of $r>1$ and then divided among the $N-1$ other players. Following the terminology of the $2$-person prisoner's dilemma, we refer to individuals that contribute to the common pool as cooperators and those that do not as defectors. Clearly, in this scenario  cooperators are doomed to extinction unless  there is some positive assortment between them \cite{Hamilton_1975,Fontanari_2024c}, which is best illustrated by the green-beard effect \cite{Hamilton_1964}, or repeated interactions that allow the emergence of reciprocity and reputation, which are not necessarily interrelated \cite{Trivers_1971,Nowak_2006} (see \cite{Wang_2023a} for a recent review). 

An obvious way to promote cooperation would be to punish defectors \cite{Boyd_1992,Boyd_2003}. In fact, this works well when it comes to eliminating defectors, but since punishment is costly, peer punishers (who also contribute to public goods) are also eliminated, leaving the resulting population of  (pure) cooperators vulnerable to invasion by defectors \cite{Boyd_1992}.  In the above scenario, a peer punisher imposes a fixed penalty on each defector. Introducing second-order peer-punishment, i.e., punishing cooperators for    not punishing defectors, leads to the elimination of both defectors and cooperators.  However, since second-order peer punishment is ineffective in the absence of defectors, because it is impossible to distinguish cooperators from punishers when there is no one to punish, cooperators may take over the population due to random drift, leaving it vulnerable to invasion by defectors again.  One way out of this conundrum is to assume that the game is noncoercive, i.e., some players may abstain from participating, and to set the payoff of nonparticipants to an ad hoc value greater than the payoff received by a defector in a group of defectors, but smaller than the payoff received by a pair of cooperators \cite{Fowler_2005,Hauert_2007}. This  guarantees that a population of defectors can be invaded by nonparticipants, which in turn can be invaded by cooperators, which can be invaded by defectors again, creating a sort of  rock-paper-scissors game scenario \cite{Hannelore_2010}. We note, however, that the dramatic scenario we describe is only true for well-mixed populations \cite{Hannelore_2010,Sigmund_2010} and that considering spatial interactions such that players interact only with a fixed number of neighbors can lead to punishers becoming dominant.  In fact, this limited set of interactions makes it possible to separate cooperators from punishers, who can then compete separately with defectors.  As a result of their indirect territorial competition, cooperators are driven out, leaving only punishers \cite{Helbing_2010a,Helbing_2010b,Wang_2024,Szolnoki_2011a}.

At least from the perspective of (modern) human social organization, a more natural way to deal with defectors of all kinds is to introduce pool or institutional punishment, where each punisher contributes a fixed amount in advance to a sanctioning institution \cite{Sigmund_2010,Szolnoki_2011}.  The fact that such costs should be paid regardless of the presence or absence of offenders is what distinguishes  institutional punishment from peer punishment, since peer punishers bear the cost of punishment only in the presence of defectors. In particular, institutional punishers not only contribute $c$ to the public goods, but also advance an amount $\gamma$ to a sanctioning institution.   Defectors are fined a fixed amount $\beta$ and cooperators are fined a smaller amount $\alpha \beta$ with $ \alpha \in [0,1]$ for not contributing to the  sanctioning institution. Not participating in the game (e.g., not recognizing the right of the sanctioning institution to punish)  is not an option so we do not include voluntary optional participation in our analysis. 
The main advantage of institutional punishment is that cooperators are easy to identify and punish even in the absence of defectors (they do not contribute to the sanctioning institution), which is not the case in the peer punishment scenario. 
This mimics the way contemporary societies are organized \cite{Ostrom_1990}.

Given the undeniably important role that institutional punishment plays in curbing defection in the real world, here we revisit some refinements to the  the standard $N$-person prisoner's dilemma formulation with pool punishment in a well-mixed population \cite{Sigmund_2010}.  In particular, we eliminate overpunishment, i.e., penalties for different types of violations are fixed and do not depend on the wealth of the sanctioning institution, which is of course proportional to the number of punishers (see \cite{Dercole_2013} for a similar approach in the context of peer punishment).  Moreover, to account for the high cost of creating and maintaining a sanctioning institution  we require a minimum number of punishers to establish it (see \cite{Vasconcelos_2013,Gois_2019} for a similar approach in the context of collective risk social dilemmas).  In addition, to reduce the burden on punishers, we assume that once this minimum number is reached, the cost of maintenance will be shared among them (see \cite{Dercole_2013,Boyd_2010} for a similar approach in the context of peer punishment).  

Here we consider three main research questions about institutional  punishment in public goods games, the answers to which are our original contributions to this widely studied topic.   First, we determine the effect of sharing the maintenance costs of the sanctioning institution on the stability of punishers. Second, in contrast to most studies of public goods games that consider small groups of players, we focus on large groups (i.e., communities) for which sanctioning institutions are ubiquitous. In analyzing the effect of group size $N$, we assume that the minimum number of punishers required to create the sanctioning institution increases linearly with $N$. Third, and more importantly, by simulating populations of finite size $M$, we study the effect of demographic noise on the predictions of the deterministic theory.

In the infinite population limit, where the game can be studied using the replicator equation framework \cite{Hofbauer_1998},  we find that the all-defectors solution is always stable and the all-cooperators solution is always unstable.  An equilibrium solution where the population consists only of institutional punishers is stable if the per capita cost to punishers of maintaining the sanctioning institution (the maximum per capita cost is $\gamma$) is less than the fine to cooperators ($\alpha \beta$) and this cost plus the cost $c$  of contributing to public goods is less than the fine to defectors ($\beta$). In this bistability scenario, there is a lower value of the frequency of punishers in the population at large that guarantees the maintenance of the sanctioning institution and, consequently, the disappearance of both types of offenders (i.e., cooperators and defectors). This threshold  has a non-trivial dependence on the size of the community.  
In addition, we find that cost sharing increases the region of stability of the all-punishers equilibrium solution as well as the size of its domain of attraction. 

For finite populations, the  game is studied through simulations of  the  imitation stochastic dynamics,  which reproduces the results of the  replicator equation framework  in the infinite population limit  \cite{Fontanari_2024b,Traulsen_2005,Sandholm_2010}. In the case the  penalty $\alpha \beta$  to cooperators is too small, so they  win over the punishers in the deterministic regime, we use finite-size scaling  \cite{Binder_1985,Privman_1990,Campos_1999} to show  that demographic noise helps to fixate cooperators even for infinitely large populations, provided, somewhat counterintuitively, that the initial fraction of cooperators vanishes with some small power of the population size $M$. Finite-size scaling is a concept from statistical physics that refers to the study of how the properties of a system change as a function of its finite size, especially near a threshold. The goal is to obtain the scaling forms and the exponents of the power-laws  that govern the behavior of quantities of interest (e.g., the  fixation probability and the mean fixation time) near a threshold, simulating large but finite system sizes.  In this sense, it provides a systematic way to computationally study the effect of demographic noise in problems that do not lend themselves to analytical approaches such as Kimura's diffusion approximation \cite{Kimura_1964}. Finite-size scaling requires simulating the system for various sizes and rescaling the variables to obtain quantities that are independent of system size, resulting in the collapse of the data for the various system sizes into a single curve, the scaling function \cite{Privman_1990}. Besides studying the effect of demographic noise, we also use this technique to study the effect of group size $N$, since the deterministic theory  gives closed analytical results for $N \to \infty$ but not for finite $N$.

The remaining sections are organized as follows.   In Section \ref{sec:model}, we present a variant of the $N$-person prisoner's dilemma that more realistically models both the maintenance and the operation of  sanctioning institutions. In particular, in this section we present the instantaneous payoffs of the three strategies that players can adopt, namely cooperators, defectors, and institutional punishers, and describe the stochastic imitation dynamics that govern the evolution of strategy frequencies for finite populations.   In Section \ref{sec:rep} we study the game for infinitely large populations using the replicator equations framework: we find the equilibrium solutions and the conditions for local stability, and explore the dynamics by numerically solving the replicator equations and representing the orbits in the simplex.    
In Section \ref{sec:stoc} we study the finite population version of the game, using Monte Carlo simulations to implement the stochastic imitation dynamics and finite-size scaling techniques to infer fixation probabilities and mean time to fixation in the limit of very large populations. In Section \ref{sec:conc}, we recapitulate our main findings and present some closing observations.

\section{The  model}\label{sec:model}

Consider a well-mixed population of finite size $M$, composed of $X$ cooperators, $Y$ defectors, and $Z$ institutional punishers, such that $X+Y+Z=M$.  Since we are only considering institutional punishers here, we will refer to these individuals simply as punishers.
A focal individual $i$ is randomly chosen at each time step $\delta t$. To determine her payoff $f_i$, we form her play group by randomly choosing  another  $N -1\geq1$  individuals  from the remaining $M -1$ individuals in the population without replacement. 

The focal individual's payoff depends on her own strategy as well as on the strategies of the other members of her group. Let us consider each possibility separately. 
 \begin{enumerate}[label=(\alph*)]
\item Assume that the focal individual  $i$ is a cooperator and her group consists of $I$ other cooperators,  $J$
defectors, and $K$ punishers.  Here $I,J,K=0, \ldots,N-1$ such that $I+J+K=N-1$.  The (instantaneous) payoff of the focal individual is
$f_i = F_C (I+1,J,K) $,  with
\begin{equation}\label{fC}
 F_C (I+1,J,K) =  \frac{(I + K)rc}{N-1} - c - \alpha \beta \Theta (K -K_m) ,
\end{equation}
where the Heaviside function is $\Theta (x) = 1$ if $x \geq 0$ and $0$ otherwise.  Here we consider the so-called others-only scenario, in which the amplified contribution $r c$  is shared by all other participants in the play group, but not by the contributor \cite{Hannelore_2010} (see also \cite{Hamilton_1975}).  This is the reason why the numerator of the first term on the rhs of Eq.\ (\ref{fC}) is missing one contributor (the total number of contributors is $I+1+K$). Moreover, each amplified contribution is shared only by $N-1$ members of the group, since the contributors do not benefit from their own contributions. Hence the factor $N-1$ in the denominator of this term. The second term on the rhs of Eq.\ (\ref{fC}) is the focal individual's contribution to public goods, and the third term is the penalty for not contributing to the sanctioning institution.  This penalty is a fraction $\alpha \in [0,1]$ of the penalty $\beta$ applied to defectors (see below) and it is only effective if there are at least  $K_m < N$  individuals  (i.e., punishers) contributing to the sanctioning institution. Once this minimum number of punishers is reached, the penalty term becomes constant, in contrast to previous formulations of institutional punishment where the penalty increases linearly with the number of individuals contributing to the sanctioning institution \cite{Hannelore_2010,Sigmund_2010}, resulting in overpunishment. In fact, the constant punishment term with $K_m=1$ was proposed in the original study that showed the non-necessity of overpunishment for the establishment of public goods   in the peer-punishment scenario \cite{Dercole_2013}.

\item Assume that the focal individual  $i$ is a defector and her group consists of $I$  cooperators,  $J$ other
defectors, and $K$ punishers so that  $I+J+K=N-1$ with $I,J,K=0, \ldots,N-1$.  The (instantaneous) payoff of the focal individual is
$f_i = F_D (I,J+1,K) $,  with
\begin{equation}\label{fD}
 F_D (I,J+1,K) =  \frac{(I + K)rc}{N-1}  - \beta \Theta (K -K_m) .
\end{equation}
The terms on the rhs of this equation have a similar interpretation to that  given in the previous item. We just note that the penalty for defectors is  a factor $1/\alpha$ higher than the penalty for cooperators when  $K \geq K_m$.
\item Assume that the focal individual  $i$ is a punisher and her group consists of $I$  cooperators,  $J$ 
defectors, and $K$ other punishers.  As before,  $I,J,K=0, \ldots,N-1$ such that $I+J+K=N-1$.  The (instantaneous) payoff of the focal individual is
$f_i = F_P (I,J,K+1) $,  with
\begin{equation}\label{fP}
 F_P (I,J, K+1) =  \frac{(I + K)rc}{N-1} -c  - \gamma \left [ 1 -   \left ( 1 -  \frac{K_m}{K+1} \right )  \Theta ( K+1-K_m) \right ].
\end{equation}
Only the last term in the rhs of this equation needs explanation. 
There is a minimum cost of creating the sanctioning institution, which can only be covered by the contribution of $K_m$ punishers.  This is similar to the variant of the NPD where a minimum number of cooperators is needed to produce the public goods \cite{Pacheco_2009}, and is also a key feature of collective risk social dilemmas \cite{Vasconcelos_2013,Gois_2019}. Once this threshold cost is reached, the   punishers optimize their payoffs by sharing among them the costs of maintaining the sanctioning institution, a feature that makes this variant of the  NPD  similar to the $N$-person snowdrift game \cite{Zheng_2007,Santos_2012,Fontanari_2024a}, and which has also been considered in the context of peer punishment \cite{Dercole_2013,Boyd_2010}.
It is interesting to mention a perhaps more realistic scenario where  punishment is graduated or adaptive, growing with the incidence of defectors and with the harm  resulting from  failure to cooperate (see, e.g., \cite{Couto_2020,Shimao_2013,Perc_2012b}). 
Another interesting scenario  considers that  incentives (bonuses to cooperators) and punishments (fines to defectors) are given externally at no cost to the players, and the problem is to optimize the incentive and punishment schedules and intensities to achieve cooperation at the lowest cost \cite{Wang_2019,Sun_2021,Wang_2023}.

To avoid the paradoxical situation that occurs for small $\gamma$, where the cost of setting up the sanctioning institution is negligible, but $K_m$ contributors are still needed to punish offenders, we could choose the per capita cost of maintaining the sanctioning institution when  the number of punishers is less than $K_m$ to be greater than or equal to the contribution to public goods, i.e., $\gamma \geq c$.  Although this is a reasonable and defensible assumption, we will allow $\gamma$ to increase from $0$ for a better visualization of our results.

\end{enumerate}

Once the payoff of the focal individual $f_i$ has been determined, we need to select a model individual $j \neq i $ in order to implement the imitation process. The model individual and her play group are  selected randomly and the payoff $f_{j}$ is determined as done for the focal individual.  The key component of the imitation dynamics is that individuals will only imitate their more successful peers, which means that the focal individual  will not change her strategy in the case that $f_{j} \leq f_{i}$.  However, when  $f_{j} > f_{i}$,  the  probability that focal individual $i$  switches to the strategy of the model individual $j$  is
\begin{equation}\label{prob0}
 \frac{f_{j} - f_{i}}{\Delta f_{\max}}, 
\end{equation}
where $\Delta f_{\max} $ is chosen so as to guarantee that this probability is not greater than 1.  In our variant of the NPD, we have 
\begin{equation}\label{Dmax}
\Delta f_{\max} = (r+1)c + \gamma ,
\end{equation}
 which is  the difference between the payoff obtained by a single defector in a group of cooperators (i.e., $rc$) and the  payoff  obtained by a single punisher in a group of defectors (i.e., $-c-\gamma$). 
 The update (or not) of the focal individual concludes the time step $\delta t$ and the time variable $t$ is increased accordingly.  The particular choice of switching probability given in Eq. (\ref{prob0}) produces the replicator equation in the $M \to \infty$ limit \cite{Fontanari_2024b} (see \cite{Traulsen_2005,Sandholm_2010} for other switching probabilities that produce the replicator equation in the large-population limit), provided that the time step is set to $\delta t = 1/M$. 

%
\section{The replicator equation formulation}\label{sec:rep}
%

Here we consider an infinitely large  population (i.e., $M \to \infty$) consisting of a fraction $x$ of cooperators, a fraction $y$ of defectors and a fraction $z$ of punishers, with $x+y+z=1$. Only the expected payoffs go into the replicator equation \cite{Hofbauer_1998}, so we need to calculate them for each of the three strategies. In particular, the expected payoff  of a cooperator is 
\begin{eqnarray}\label{piC}
\pi^C   &  = &     \sum_{I,J,K=0}^{N-1} \frac{(N-1)!}{I!J!K!} x^I y^J z^K F_C (I+1,J,K)  \delta_{I+J+K,N-1} \nonumber \\
&  = &  rc(x+z) - c - \alpha \beta \left [ 1 - \sum_{K=0}^{K_m-1}\binom{N-1}{K}  z^K  (1-z)^{N-1-K}  \right ] , 
\end{eqnarray}
and  the expected payoff of a defector is
\begin{eqnarray}\label{piD}
\pi^D   &  = &   \sum_{I,J,K=0}^{N-1} \frac{(N-1)!}{I!J!K!} x^I y^J z^K  F_D (I,J+1,K)  \delta_{I+J+K,N-1} \nonumber \\
&  = &  rc(x+z)  - \beta \left [ 1 - \sum_{K=0}^{K_m-1}\binom{N-1}{K} z^K  (1-z)^{N-1-K}  \right ] .
\end{eqnarray}
Calculating the expected payoff of a punisher is a bit more complicated  and yields
\begin{eqnarray}\label{piP}
\pi^P   &  = &   \sum_{I,J,K=0}^{N-1} \frac{(N-1)!}{I!J!K!} x^I y^J z^K  F_P (I,J,K+1)  \delta_{I+J+K,N-1} \nonumber \\
&  = &  rc(x+z)  - c  - \frac{\gamma K_m}{Nz}  \left [ 1 - \sum_{K=0}^{K_m-1}\binom{N}{K} z^K  (1-z)^{N-K}  \right ] \nonumber \\ 
&  &  - \gamma \sum_{K=0}^{K_m-2}\binom{N-1}{K}  z^K  (1-z)^{N-1-K} ,
\end{eqnarray}
where it is implicit that the last term in the rhs  is zero if $K_m =1$. 

Sanctioning institutions only make sense for very large play groups, otherwise peer punishment or small coalitions are more appropriate \cite{Ostrom_1990}. It is therefore of interest to obtain the expressions for the payoffs of the strategies in the limits $N \to \infty$ and $K_m \to \infty$, such that the ratio 
\begin{equation}
\zeta = \frac{K_m}{N} 
\end{equation}
is nonzero. This can be achieved by replacing the binomial distributions by Gaussian distributions in Eqs. (\ref{piC}), (\ref{piD}) and (\ref{piP}) and doing the integrations over the number of punishers  (see \ref{ref:A}). The final result is 
\begin{equation}\label{piCinf}
\pi^C=
    \begin{cases}
      rc(x+z) - c  & \text{if $z < \zeta $ }\\
        rc(x+z) - c  -\alpha \beta/2 & \text{if $z =\zeta $ } \\
      rc(x+z) - c  -\alpha \beta & \text{if $z > \zeta $},
    \end{cases}       
\end{equation}
\begin{equation}\label{piDinf}
\pi^D=
    \begin{cases}
      rc(x+z)   & \text{if $z < \zeta $ }\\
        rc(x+z)   - \beta/2 & \text{if $z =\zeta $ } \\
      rc(x+z)   - \beta & \text{if $z > \zeta $},
    \end{cases}       
\end{equation}
and
\begin{equation}\label{piPinf}
\pi^P=
    \begin{cases}
      rc(x+z) -c  - \gamma & \text{if $z \leq \zeta $ }\\
      rc(x+z)   -c  - \gamma \zeta/z & \text{if $z > \zeta $}.
    \end{cases}       
\end{equation}

 The replicator equations governing the evolution of the frequency  cooperators, defectors and punishers  in the infinite population are
\begin{eqnarray}
\frac{dx}{dt}  & = &  x   \left [ \pi^C (x,z) - \bar{\pi} (x,y,z)  \right ]  \frac{1}{\Delta f_{\max}}  \\
\frac{dy}{dt}  & = &y   \left [ \pi^D (x,z) - \bar{\pi} (x,y,z)  \right ]  \frac{1}{\Delta f_{\max}}   \\
\frac{dz}{dt}  & = &z   \left [ \pi^P (x,z) - \bar{\pi} (x,y,z)  \right ]  \frac{1}{\Delta f_{\max}}  ,
\end{eqnarray}
where $\bar{\pi} = x \pi^C + y \pi^D + z \pi^P$ is the average payoff of the population and $\Delta f_{\max} $ is given by Eq.\ (\ref{Dmax}). Although we could eliminate the factor $1/\Delta f_{\max}$ by a proper rescaling of $t$ and thus obtain the standard form of the replicator equation, keeping it facilitates comparison with the finite population simulations presented in Section \ref{sec:stoc}. As already mentioned, these equations are derived as the infinite population limit of the imitation dynamics described  in Section \ref{sec:model} (see \cite{Fontanari_2024b}).
  Since $x+y+z=1$ and the expected payoffs of the different strategies depend only on $x$ and $z$, henceforth we will eliminate $y$ in favor of these  variables.  The resulting equations become 

\begin{eqnarray}
\frac{dx}{dt}  & = & x   \left [ (1-x) (\pi^C  - \pi^D)  - z(\pi^P-\pi^D)   \right ]  \frac{1}{\Delta f_{\max}}  \label{x}\\
\frac{dz}{dt}  & = & z   \left [  (1-z) (\pi^P  - \pi^D)  -x(\pi^C-\pi^D)    \right ] \frac{1}{\Delta f_{\max}}  \label{z} ,
\end{eqnarray}
where we have omitted the dependence of the expected payoffs of the strategies on $x$ and $z$ for simplicity.   Since only  payoff differences 
appear in these equations,  we can drop  the common  term $rc(x+z)$ in Eqs. (\ref{piC}), (\ref{piD}) and  (\ref{piP}), which greatly facilitates our analysis of the equilibrium solutions.   On the one hand, as $z$ increases from $0$ to $1$,  $\tilde{\pi}^C(z) =\pi^C(x,z) - rc(x+z)$    monotonically decreases  from $-c$  to $-c - \alpha \beta$. Similarly,  $\tilde{\pi}^D(z) =\pi^D(x,z) - rc(x+z)$  monotonically decreases  from  $0$  to $- \beta$.
On the other hand,  $\tilde{\pi}^P(z) =\pi^P(x,z) - rc(x+z)$ monotonically  increases from $-c-\gamma$ to $-c - \gamma \zeta$,  as $z$ increases from $0$ to $1$.  The introduction of these shifted payoffs makes it clear that   the  equilibrium strategy frequencies do not depend on  the amplification factor $r$.

Since the expected payoffs $\pi^C$, $\pi^D$, $\pi^P$, and the maximum instantaneous payoff difference $\Delta f_{\max}$ are measured in the same payoff units, the time $t$ in the replicator equations (\ref{x}) and (\ref{z}) is dimensionless.  This is expected because time in the imitation stochastic dynamics is measured in Monte Carlo steps, where a Monte Carlo step corresponds to the choice of  $M$ random  focal individuals for the update.  Finally, we note that by measuring all parameters in units of the contribution $c$ to the public goods, we can set 
$c=1$ in the numerical analysis without loss of generality.

The fixed-point or equilibrium solutions  of the replicator equations  (\ref{x}) and    (\ref{z}), denoted by  $x^*$ and $z^*$, are obtained by setting $dx/dt=dz/dt=0$. In the following we briefly present these solutions together with the results of a standard  local stability analysis. 
This analysis is done by linearizing the replicator equations at the equilibrium solutions, resulting in the linear system 
\begin{equation}
\begin{pmatrix} 
\frac{du}{dt} \\ 
 \frac{dv}{dt}  
\end{pmatrix} = \mathbf{A}
\begin{pmatrix} 
u \\ 
 v   
 \end{pmatrix} ,
\end{equation}
where $u=x-x^*$ and $v=z-z^*$. The $2 \times 2$ Jacobian matrix  $ \mathbf {A} $  evaluated at the equilibrium solutions is called the community matrix and the signs of its eigenvalues determine the local stability of these solutions \cite{Britton_2003,Murray_2007}.

\subsection{All-cooperators solution}

 Clearly,  $x=1$ and consequently $y=z=0$ is a solution of  $dx/dt=dz/dt=0$ and corresponds to  a population consisting only of cooperators.    This solution is locally stable provided that the eigenvalues $\lambda_1$ and $\lambda_2$ of the $2 \times 2$ community matrix are negative.   They are
 \begin{eqnarray}
 \lambda_1 & = & c \\
 \lambda_2 & = &  - \gamma .
 \end{eqnarray}
 Since $\lambda_1 > 0$ and $\lambda_2 <0$, the all-cooperators solution  is a saddle point. This means that a population of cooperators can be invaded by defectors, but not by punishers. The population mean payoff $\bar{\pi} = c(r-1)$ is the largest possible payoff for a population at equilibrium.

\subsection{All-defectors solution}

In this case  $y=1$ and consequently $x=z=0$. This is clearly a solution of  $dx/dt=dz/dt=0$. The eigenvalues of the corresponding community matrix are
 \begin{eqnarray}
 \lambda_1 & = &- c \\
 \lambda_2 & = & -c  - \gamma,
 \end{eqnarray}
which are always negative and so a population of defectors  cannot be invaded, i.e., the all-defectors solution is an  evolutionarily stable equilibrium \cite{Maynard_1982}. However, the population mean payoff $\bar{\pi} = 0$ is the worst possible payoff for a population in equilibrium, which reveals the origin of the public goods dilemma: although the highest individual payoff is obtained by a defector in a group of cooperators ($rc$), if every player decides to defect, the population as a whole ends up in the worst possible situation.

\subsection{All-punishers solution}

In this case  $z=1$ and consequently $x=y=0$.  As in the previous cases, this is clearly a solution of  $dx/dt=dz/dt=0$ and  the eigenvalues of the corresponding community matrix are
 \begin{eqnarray}
\lambda_1 &  = &  -\alpha \beta + \gamma \zeta  \label{lpu1} \\
\lambda_2 & = &  \gamma \zeta + c - \beta .  \label{lpu2}
 \end{eqnarray}
Now things get more interesting. The stability condition $\lambda_1 < 0$ means that the per capita cost to punishers of contributing to the sanctioning institution must be less than the penalty for cooperators, while $\lambda_2 < 0$ means that the cost of contributing to public goods and to the sanctioning institution must be less than the penalty for defectors. These are reasonable conditions that are, in principle, likely to be satisfied in real societies that use sanctioning institutions to deter offenders.
 
 \subsection{No-cooperators solution}
 
\begin{figure}[th] 
\centering
 \includegraphics[width=1\columnwidth]{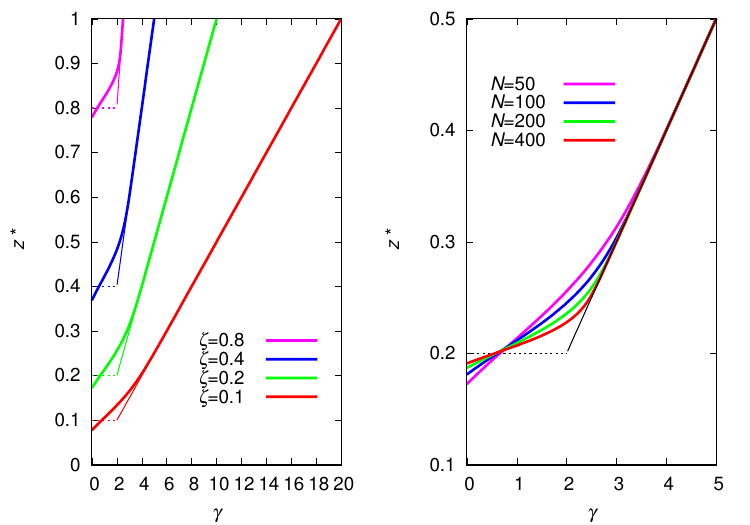}  
\caption{Unstable fixed point $z^*$  corresponding to the no-cooperators solution as a function of the maximum per capita contribution to the sanctioning institution $\gamma$. Left panel:  $N=50$ and  $\zeta = 0.1, 0.2, 0.4, 0.8$, as indicated. Right panel: $\zeta = 0.2$ and  $N=50,100, 200, 400$, as indicated.
The  thick curves are the results for finite $N$, while the  thin  lines are the results for $N \to \infty$, which exist only  for $z^* \geq \zeta$. The other  parameters are $\beta =3$ and $c=1$.
 }  
\label{fig:1}  
\end{figure}

The scenario with no cooperators, i.e.  $x=0$ and $y= 1-z^*$, is a solution of $dx/dt=dz/dt=0$, provided that $z^*  \in (0,1)$ is a root of $g (z) = \tilde{\pi}^D(z) -\tilde{\pi}^P (z) $.  Since $g(z)$ decreases  monotonically from $c+\gamma $ to $-\beta + c + \gamma \zeta $ as $z$ increases from $0$ to $1$, it is clear that the single valid root exists only if  $-\beta + c + \gamma \zeta < 0$, which is  the  only condition for the stability of the all-punishers fixed point  in the absence of cooperators. Therefore, the no-cooperators solution is unstable and the value of $z^*$ delimits the basins of attraction of the all-punishers ($z=1$)  and the all-defectors ($y=1$)  solutions.  More precisely, if $z(0) > z^*$  the dynamics is driven to the all-punishers fixed point and to the all-defectors fixed point if $z(0) < z^*$.  So the larger $z^*$ is, the smaller is the size of the domain of attraction of  the all-punishers solution. Figure \ref{fig:1} shows $z^*$  as function of $\gamma$ for different values of the ratio $\zeta$. As expected, the  domain of attraction of the  all-punishers fixed point decreases with increasing $\gamma$ and $\zeta$.
Note that $z^*$ does not depend on $\alpha$  since the cooperators play no role in this scenario.

While for finite $N$ we have to obtain the root $z^*$ of $g(z)$ numerically, for $N \to \infty$ it can be obtained easily using Eqs.\ (\ref{piDinf}) and (\ref{piPinf}), leading to
\begin{equation}\label{znocinf}
z^* = \frac{ \gamma \zeta}{\beta - c}
\end{equation}
provided that  $ \gamma > \beta - c $, which ensures that $z^* > \zeta$ and so that the offenders are punished. 
For  $ \gamma < \beta - c $ there is no root for $g(z)$, which means that  the  size of the domain of attraction of the all-punishers solution vanishes. 
For $ z^* = \zeta $, equating Eqs.\ (\ref{piDinf}) and (\ref{piPinf}) yields $\gamma = \beta/2 - c$.   Interestingly, the curves for different $N$ in the right panel of Fig.\ \ref{fig:1} intersect at the coordinates $\gamma = \beta/2 - c = 0.5$ and $ z^* = \zeta =0.2$, signaling a threshold (i.e., a discontinuity) at $z^*=\zeta$ when $\gamma $ is considered as a function of $z^*$, rather than the reverse. The results show, as expected, the strong effect of demographic noise near the threshold.  It is easy to understand the dependence on $N$ shown in the right panel of Fig.\ \ref{fig:1}.  On the one hand, for  $z^* < \zeta$  the fluctuations in the composition of small groups allow a fair probability that $K \geq K_m$ in some groups (and so the defectors are punished), guaranteeing the existence of the  no-cooperators solution  in the region $ z^* = \zeta $ for finite $N$.  For a fixed $\gamma$, the smaller the group size, the lower the value of $z^*$ needed to guarantee  the existence of this solution, as shown in the figure.  On the other hand, if $z^* > \zeta$, the same fluctuations allow defectors to go unpunished in some small groups, so for a fixed $\gamma$, the smaller the group size, the higher the value of $z^*$ needed to guarantee the existence of the no-cooperators solution. This is the reason why the effect of increasing $N$ is  reversed  depending on whether $z^*$ is greater or less than $\zeta$. In addition, variation of $N$ has no effect at all if $z^* = \zeta$.

  \subsection{No-defectors solution}

\begin{figure}[th] 
\centering
 \includegraphics[width=1\columnwidth]{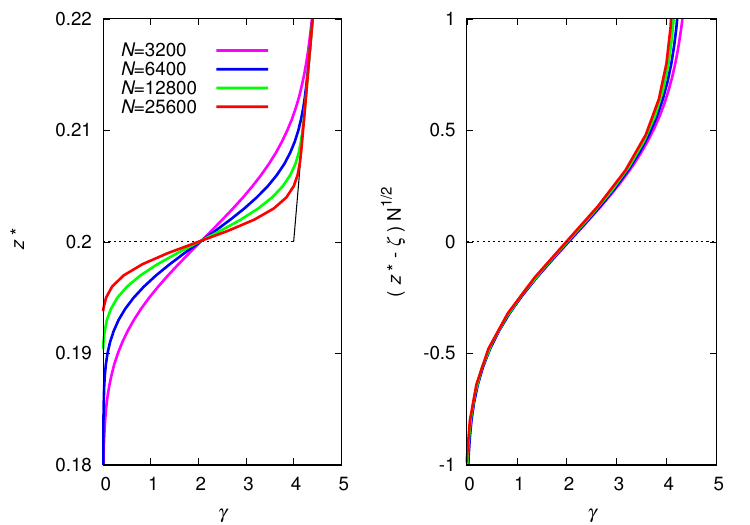}  
\caption{The no-defectors unstable fixed point $z^*$ as a function of the maximum per capita contribution to the sanctioning institution $\gamma$ for  $\zeta=0.2$, $\alpha \beta =4$, and  $N = 3200, 6400, 12800, 25600$, as indicated (left panel). The black thin line is the result for $N \to \infty$.   The right panel shows the same curves with the y-axis rescaled to $(z^* - \zeta)N^{1/2}$. The color convention is the same for both panels.
 }  
\label{fig:2}  
\end{figure}

The analysis of the scenario without defectors  is very similar to that presented above. In particular,   $y=0$ and $x= 1-z^*$ is
a solution of $dx/dt=dz/dt=0$, provided that $z^*  \in (0,1)$ is a root of $h (z) = \tilde{\pi}^C(z) -\tilde{\pi}^P (z) $.  Using the same reasoning as before, we conclude that  the single valid root exists only if  $-\alpha \beta  + \gamma \zeta< 0$, which is  the  only condition for the stability of the all-punishers fixed point  in the absence of defectors. We recall that in this case the all-cooperators fixed point is stable.  Therefore, the no-defectors solution is unstable and the value of $z^*$ delimits the basins of attraction of the all-punishers ($z=1$)  and the all-cooperators ($x=1$)  solutions.
As before, $z^*$ can be obtained easily using Eqs. (\ref{piCinf}) and (\ref{piPinf}) for $N \to \infty$, leading to
\begin{equation}\label{znodinf}
z^* = \frac{  \gamma \zeta}{\alpha \beta}     
\end{equation}
provided that $ \gamma > \beta \alpha  $ which ensures that $z^* > \zeta$.  For $z^* = \zeta$ we have $\gamma = \alpha \beta/2$. 
 Of course, the results for both the finite and infinite play group sizes   depend  on the product $\alpha \beta$ and not on the individual values of the parameters $\alpha $ and $\beta$. In addition, the parameter $c$  plays no role in the determination of  $z^*$ for the no-defectors solution. Since the dependence of $z^*$ on $\gamma$ is very similar to that shown in Fig.\ \ref{fig:1}, here we look at how the  finite  play group solutions approach the 
discontinuous  result predicted by the analytical solution  in the limit $N \to \infty$.
The left panel of Fig.\  \ref{fig:2} shows the unstable fixed point $z^*$ for fixed $\zeta = 0.2$ and $\alpha \beta = 4$,  and a variety of large play group sizes $N$.   The right panel shows that the deviation from the $N \to \infty$ prediction vanishes like $1/\sqrt{N}$ as $N$ increases.   This is expected since for large $N$ we have used the approximation (see \ref{ref:A})
\begin{equation}\label{erf1}
\sum_{K=0}^{K_m-1}\binom{N}{K} z^K  (1-z)^{N-K} \approx  \frac{1}{2} \mbox{erf} \left [  \frac{ (\zeta -z) \sqrt{N}}{\sqrt{2z(1-z)}}  \right ]
+
\frac{1}{2} \mbox{erf} \left [  \frac{ z \sqrt{N} }{\sqrt{2z(1-z)}} \right ]
\end{equation}
to derive Eqs.\ (\ref{znocinf}) and  (\ref{znodinf}).  We note that finding how the deviations scale with $N$ by `collapsing'
the curves for different values of $N$  into a single curve is a technique widely used in finite-size scaling analysis to calculate the critical exponents of phase transitions (threshold phenomena) \cite{Binder_1985,Privman_1990,Campos_1999}. In fact, there is a threshold at $z^* = \zeta$, indicated by the crossing of the curves for different $N$ at $\gamma = \alpha\beta/2 $ and $z^* = \zeta $.  As before, the effect of increasing $N$ is reversed depending on whether $z^*$ is greater or less than $\zeta$. For $N \to \infty$, the punishers can be maintained in the population only if $z^* > \zeta$, but for finite $N$ there is a fair probability that $K \geq K_m$ in some groups even for $z^* < \zeta$, which is enough to maintain the  punishers in the population. 

\begin{figure}[th] 
\centering
 \includegraphics[width=1\columnwidth]{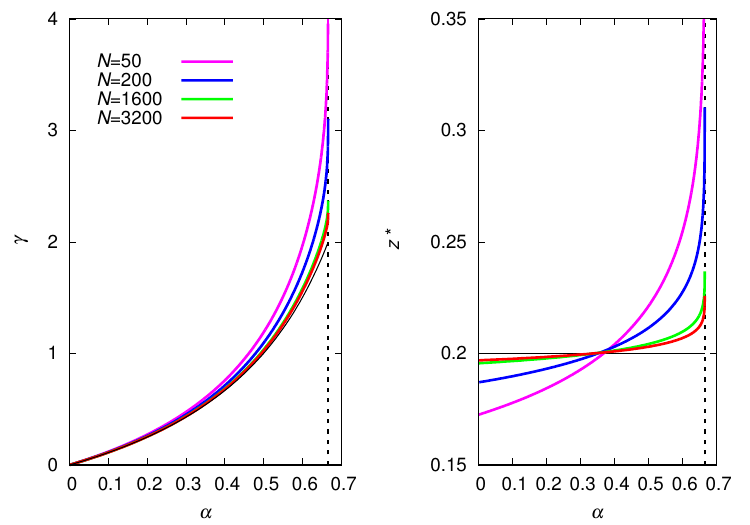}  
\caption{Values of $\alpha$ and $\gamma$ (left panel) for which the coexistence fixed point $z^*$  (right panel) exists for  $\zeta=0.2$,  and  $N = 50, 200, 1600, 3200$, as indicated. The color convention is the same for both panels. The other parameters are $\beta=3$ and $c=1$. The vertical dashed lines indicate the threshold $\alpha = 1-c/\beta = 2/3$ beyond which there is no coexistence solution. The black thin curves indicate the results for $N \to \infty$.
}  
\label{fig:3}  
\end{figure}

\subsection{Coexistence solution}

The existence of an equilibrium solution where all three strategies are present would require a value of $z^*$ such that $g (z^*) = \tilde{\pi}^D(z^*) -\tilde{\pi}^P (z^*) =0 $ and  $h (z^*) = \tilde{\pi}^C(z^*) -\tilde{\pi}^P (z^*) =0 $.  Solving these equations simultaneously, we find that such values do indeed exist for certain values of the model parameters, as shown in Fig.\ \ref{fig:3}. However, the above equilibrium conditions only determine the value of $z^*$, so $x^*$ must depend on the initial conditions. 
In fact, if we set $z(0) = z^*$, then the replicator dynamics freezes at $x^* = x(0)$, which means that there are infinitely many coexistence fixed points.  

The derivation of the results for  $N \to \infty$ is a bit more complicated than in the previous cases. The reason is that as $N$ increases, $z^*$ approaches $\zeta$ so that $( \zeta -z^*)N^{1/2} = \eta$ is finite and nonzero (see \ref{ref:A}). Using the approximation (\ref{erf1}) we can write the equations $g (z^*)=h (z^*) =0$ as
\begin{equation}\label{etac}
\alpha = 1 - \frac{c}{\beta [1-H(\eta)]}
\end{equation}
\begin{equation}\label{gammac}
\gamma = \frac{c \alpha}{1-\alpha}
\end{equation}
where
\begin{equation}
 H(\eta) = \frac{1}{2}  + \frac{1}{2} \mbox{erf} \left [ \frac{\eta}{\sqrt{2\zeta(1-\zeta)}} \right ] .
\end{equation}
For fixed $\beta$, the largest value of $\alpha$ for which  Eq.\ (\ref{etac})  admits a solution is $ \alpha = 1-c/\beta $,  corresponding  to $\eta \to - \infty$, and in this case we have $\gamma =  \beta -c$. The left panel of Fig.\ \ref{fig:3} shows the  very good agreement between  Eq.\ ( \ref{gammac}) and the numerical results for large $N$. Of course, whenever Eq.\ (\ref{etac}) admits a solution we have $z^* = \zeta$, which is indeed the  large $N$ limit  of the numerical results, as shown in the right panel of Fig.\ \ref{fig:3}. This panel shows that for finite $N$, somewhat counterintuitively, increasing $\alpha$ does not favor the punishers, since $z^*$ increases with $\alpha$, resulting in a reduction of the basin of attraction of the all-punishers solution. The reason is that for the coexistence solution to exist, $\gamma$ must increase with $\alpha$ (see the left panel of Fig.\ \ref{fig:3}), which is detrimental to the punishers.

 This completes the analysis of the  fixed points of the replicator equations (\ref{x}) and  (\ref{z}).  There is no fixed point corresponding to the no-punishers scenario (i.e., $z=0$ and $y = 1-x$ with $x \in (0,1)$) since  $ \tilde{\pi}^D(0) > \tilde{\pi}^C(0)$ for all parameter settings.We note that since the eigenvalues of the community matrices for the stable (i.e., all-punishers and all-defectors) fixed points do not depend on $N$, the group size only affects the sizes of the domains of attraction of these fixed points.

 \subsection{Dynamics}

The all-defectors fixed point ($y=1$) is stable for all values of the model parameters, while the all-punishers fixed point ($z=1$) is stable if the conditions $\gamma \zeta < \alpha \beta$ and $\gamma \zeta < \beta -c$ are satisfied. Except for the coexistence and for the no-cooperators fixed points, which are unstable (see figures below), all other fixed points and especially the all-cooperators fixed point ($x=1$) are saddle points. Thus, the useful information we can obtain from the study of the replicator dynamics, i.e., the numerical solution of Eqs. (\ref{x}) and (\ref{z}), is the domains of attraction of the stable fixed points, which are better visualized by ternary or simplex plots.   In these plots,  the vertices of the triangle  denote homogeneous populations of punishers (P), cooperators (C),  and  defectors (D).  As usual, we represent the   stable points by a filled (black) circle and an unstable or a saddle point by an open (white) circle.

\begin{figure}[th] 
\centering
 \includegraphics[width=1\columnwidth]{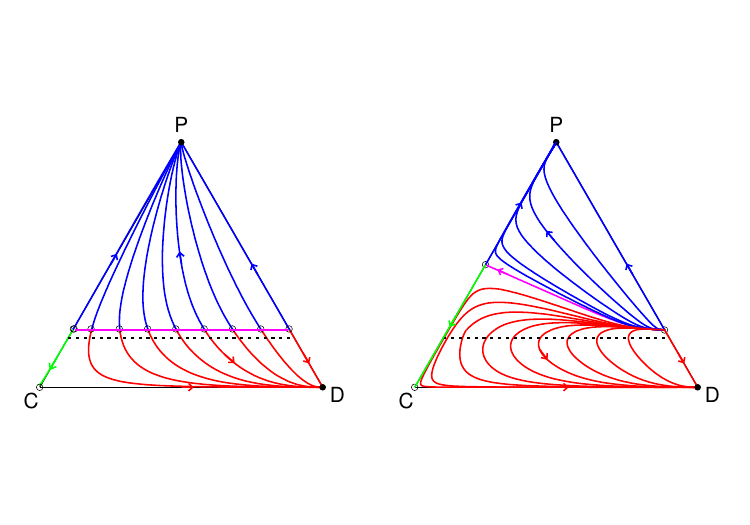}  
\caption{Replicator dynamics for (left panel) $\alpha = 0.549$ and $\gamma=1.499$ and (right panel)  $\alpha = 0.2$ and $\gamma=1.499$. The other parameters are  $N=50$,  $K_m=10$,  $\beta=3$ and $c=1$.  The filled circles indicate the stable fixed points and the open circles the saddle and  unstable fixed points.  The blue orbits converge to the all-punishers fixed point, the red orbits converge to the all-defectors fixed point, and the green orbits converge to the all-cooperators fixed point as indicated by the arrows in representative orbits. The magenta horizontal  line in the left panel is a line of (unstable) coexistence fixed points and is the separatrix delimiting the domains of attraction of P and D. The magenta curve in the right panel is the separatrix over which the dynamics goes to the no-defectors fixed point, as indicated by the arrow. The black dashed line is $z = \zeta =0.2$.  
}  
\label{fig:4}  
\end{figure}

Figure \ref{fig:4} shows the replicator dynamics in the case of bistability of the  all-punishers and all-defectors fixed points. The starting points of the orbits are very close to the unstable fixed points. The left panel illustrates the scenario where the coexistence fixed points exist (the model parameters must be set very precisely according to Fig.\ \ref{fig:3}): since the coexistence condition only fixes the frequency of punishers $z^*$, there are infinitely many unstable coexistence fixed points determined by the different values of the initial frequency of cooperators $x^* = x(0)$. Of course, the line of the unstable fixed point coincides with the separatrix delimiting the domains of attraction of the all-punishers and all-defectors fixed points. The right panel illustrates the more common bistability scenario, where there are no coexistence fixed points and the boundary of the attraction domains is determined by a separatrix, which directs the flow toward the no-defectors fixed point.  
Since the only difference between the two panels is the value of $\alpha$, the no-cooperators fixed point is the same in both plots. In fact, we can now  see that this fixed point is unstable, whereas the no-defectors fixed point is a saddle. This is only true for the scenario of bistability between the all-punishers and the all-defectors fixed point, otherwise the no-defectors fixed point is also unstable.

\begin{figure}[th] 
\centering
 \includegraphics[width=1\columnwidth]{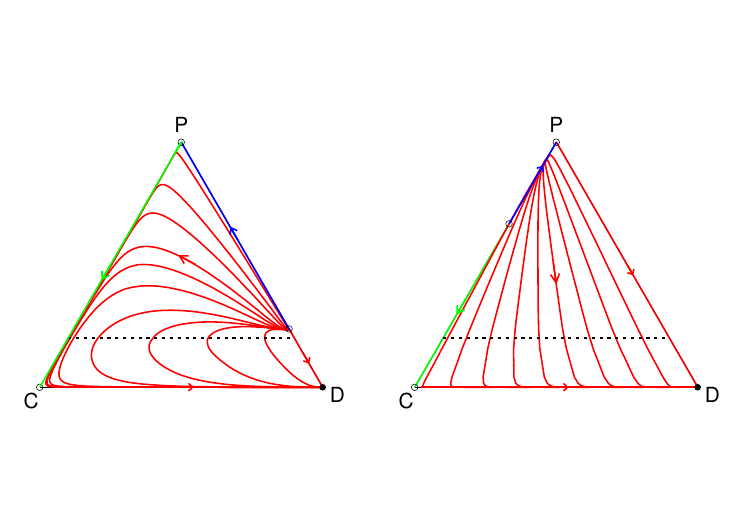}  
\caption{Replicator dynamics for (left panel) $\alpha = 0.05$ and $\gamma=1.5$ and (right panel)  $\alpha = 0.9$ and $\gamma=12.5$. The other parameters are  $N=50$,  $K_m=10$,  $\beta=3$ and $c=1$. The filled circles indicate the stable fixed points and the open circles the saddle and  unstable fixed points.  The blue orbits converge to the all-punishers fixed point, the red orbits converge to the all-defectors fixed point, and the green orbits converge to the all-cooperators fixed point as indicated by the arrows in representative orbits.  The black dashed line is $z = \zeta =0.2$.  
}  
\label{fig:5}  
\end{figure}

 Figure \ref{fig:5}  shows the scenarios where the all-punisher fixed point is a  saddle and thus the   all-defectors  fixed point is the only stable solution of the replicator dynamics. The left panel illustrates the case  $\gamma \zeta > \alpha \beta$ and $\gamma \zeta < \beta -c$ so that the cooperators win over the punishers. A typical orbit leads to an almost complete dominance of the population by cooperators, which are then quickly overrun by defectors since the small number of punishers is not enough to establish a sanctioning institution.  The right  panel illustrates the case  $\gamma \zeta < \alpha \beta$ and $\gamma \zeta > \beta -c$ so that the  defectors win over the punishers. As before, the starting points of the orbits are close to the unstable  fixed points at the edges of the simplex. The frequency of both cooperators and defectors increases at the expense of the punishers, but once they disappear, the defectors quickly take over the population.
 
\begin{figure}[th] 
\centering
 \includegraphics[width=1\columnwidth]{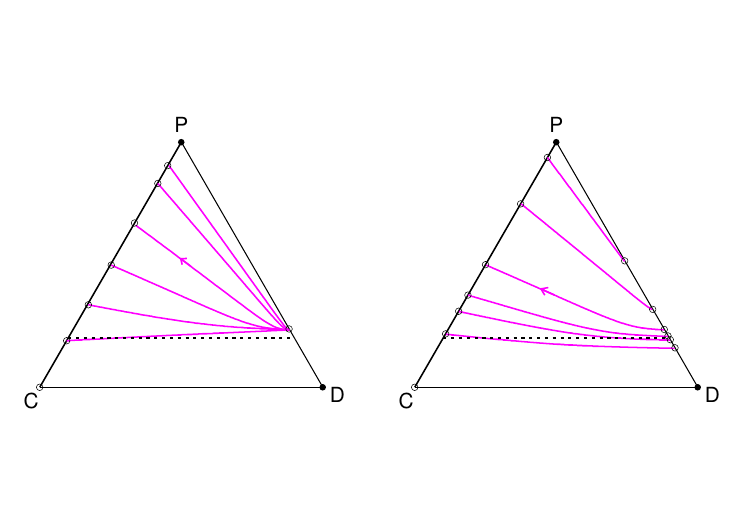}  
\caption{Separatrices  delimiting the domains of attraction of P and D for $\gamma = 1.5$ and (left panel from top to bottom) $\alpha = 0.11, 012, 0.15, 0.2, 0.3, 1$ and $\beta=3$, and  (right panel from top to bottom) $\beta =1.6, 2, 3,4,5, 10 $ and $\alpha=0.2$. The other parameters are  $N=50$,  $K_m=10$, and $c=1$. The filled circles indicate the stable fixed points and the open circles the saddle and  unstable fixed points.  
The flow over a separatrix is always towards the no-defectors fixed point, as indicated by the arrow in representative orbits.  The black dashed line is $z = \zeta =0.2$.  
}  
\label{fig:6}  
\end{figure}

 The regime  of bistability between the all-punishers and the all-defectors fixed points is the more interesting feature of the replicator dynamics. The resulting  scenario can be better appreciated by the separatrix curves that delimit the domains of attraction of these fixed points. These curves are shown in Fig. \ref{fig:6} for a variety of values of the parameters $\alpha$ (left panel) and $\beta$ (right panel) such that $\gamma \zeta < \alpha \beta$ and $\gamma \zeta < \beta -c$.  Note that $z^* = \zeta$ is a lower bound on the frequency of punishers only in the limit $N \to \infty$. For finite $N$, there is a nonzero probability of play groups with $K>K_m$ punishers even if $z < \zeta$ (see Figs.\ \ref{fig:1}, \ref{fig:2}, and \ref{fig:3}).

Figures \ref{fig:4} and \ref{fig:5}  illustrate the typical dynamic scenarios  that hold for finite and  infinite play group sizes. In particular, the  right panel of Fig. \ref{fig:4} makes clear that in  the presence of cooperators  a larger number of punishers is required  to eliminate the threat of defectors, since the no-defectors fixed point occurs at a larger value of z than the no-cooperators fixed point.  The left panel of Fig. \ref{fig:5} shows the detrimental effect of the presence of improperly punished cooperators: they exploit the rarity of defectors to overrun the punishers, opening the way for the invasion of defectors.

A word is in order about the effect of changing the group size $N$ on the orbits shown in the above simplexes. As noted earlier, changing $N$ only affects the sizes of the domains of attraction of the stable fixed points, so its effect is negligible in the scenarios shown in Fig. \ref{fig:5}, where the only stable fixed point is the all-defectors fixed point.  In the more interesting scenario of bistability between the all-defectors and all-punishers fixed points shown in Fig. \ref{fig:4}, increasing $N$ simply shifts the separatrices slightly. For example, considering the parameter setting used in the right panel of Fig. \ref{fig:4}, for $N=10$ and $K_m=2$ we find that the equilibrium frequency of punishers for the no-defectors fixed point is $z_{nd}^* = 0.506$, while for the no-cooperators fixed point it is $z_{nc}^* = 0.242$. We recall that these fixed points determine the end points of the separatrix.  For $N \to \infty$  and $\zeta = 0.2$,  we find  $z_{nd}^* = 0.5$ and $z_{nc}^* = 0.2$, so the separatrices for $N=10$ and $N \to \infty$ differ very little.

%
\section{Finite population simulations}\label{sec:stoc}
%

As already pointed out, a major advance in the theoretical study of public goods games  is the realization that the replicator equation formalism originally introduced to study biological evolution in continuous time \cite{Hofbauer_1998} also describes the social dynamics scenario where individuals are more likely to  imitate the behavior of their more successful peers   \cite{Fontanari_2024b}. In this section, we offer a comparison between the predictions of the deterministic replicator equations (\ref{x}) and (\ref{z}) and the finite population simulations  using the stochastic  imitation dynamics presented in Section \ref{sec:model}.

Although the probability that the focal individual switches strategy, Eq.\  (\ref{prob0}), depends on the difference between  her instantaneous payoff  and the payoff of the model individual, their shares of the public goods  do not cancel out because they are evaluated for  different play groups. Thus, in principle, the simulation results could depend on the amplification factor $r$. Of course, we expect this effect to be negligible for large population sizes $M$. 

\begin{figure}[t] 
\centering
 \includegraphics[width=1\columnwidth]{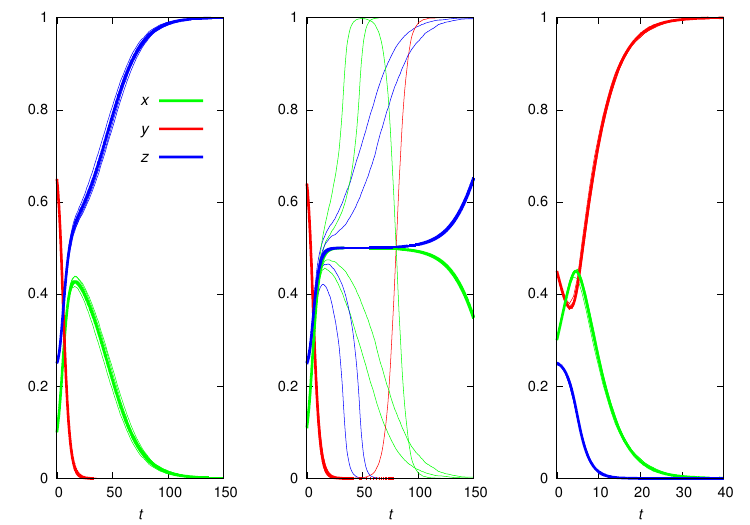}  
\caption{Frequencies of  cooperators $x$, defectors $y$ and punishers $z$  as a function of time for the  initial conditions $z(0)=0.25$ and  $x(0) = 0.1$   (left panel), $x(0) = 0.11052$  (middle panel) and   $x(0) = 0.3$ ( right panel).
The thin  curves are the four runs of the stochastic  simulation algorithm  for $M=10^4$ and the thick curves are  the numerical solutions of the replicator equations. The color convention is the same for the three panels.
 The  parameters  are $N=50$, $K_m=10$, $\alpha = 0.2$, $\gamma = 1.5$, $\beta=3$, $r=2$ and $c=1$. 
 }  
\label{fig:7}  
\end{figure}

Figure \ref{fig:7} shows  the numerical solution of  Eqs. (\ref{x}) and (\ref{z})  together with  four independent   runs of the stochastic  simulation algorithm  for  populations  of size  $M=10^4$ and the same parameters of the simplex plot  in the right panel of Fig.\ \ref{fig:4}.  The finite-size effects in the dynamics are very small  for initial conditions far from the separatrix (left and right panels). These two panels illustrate the nonintuitive result that increasing the number of cooperators while keeping the number of punishers fixed (and thus decreasing the number of defectors) leads to the dominance of defectors.   The middle panel shows the dynamics when the initial conditions $x(0)$ and $z(0)$ are very close to the separatrix. If it were possible to choose these initial values exactly at the separatrix and to evolve the deterministic dynamics without  round-off errors, then the dynamics should lead to the no-defectors fixed point, $x^*=z^* = 0.5$ for the parameter setting of the figure. In fact, if these small numerical errors eventually move the deterministic trajectory away from the  no-defectors fixed point, then we expect the effect of the intrinsic noise of the finite population simulations to be much more disruptive.  This is indeed the case: among the four runs of the stochastic algorithm, two lead to the  fixation of the punishers, one to the fixation of the cooperators, and one to the  fixation of the defectors. Most interestingly, since  the orbits leading to the  all-defectors fixed point may be too close to the no-defectors edge of the simplex (see the right panel of Fig.\ \ref{fig:4} and the left panel of  Fig.\ \ref{fig:5}), the demographic noise is likely to favor the fixation of the cooperators.  

\begin{figure}[th] 
\centering
 \includegraphics[width=1\columnwidth]{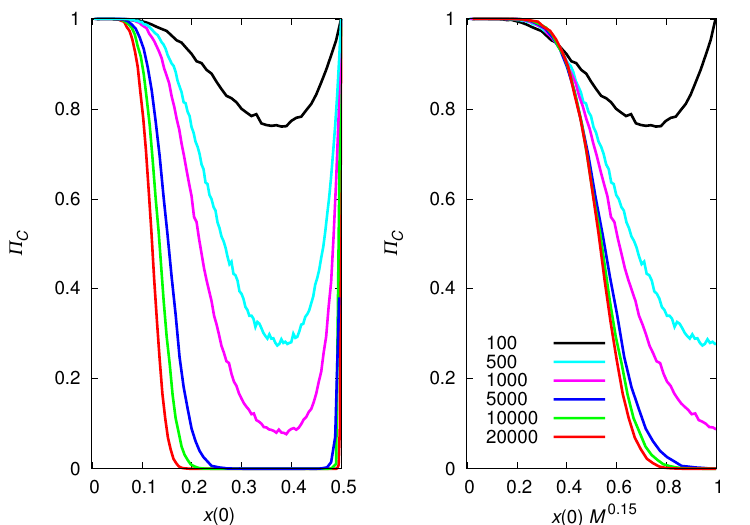}  
\caption{Probability of fixation of cooperators $\Pi_C$ as a function of their initial frequency (left panel) for populations of sizes (from top to bottom) $M=100, 500, 1000, 5000, 10000$ and $20000$. The right panel shows  $\Pi_C$ as a function of the scaled variable $x(0)  M^{0.15}$. The color convention is the same for both panels. The initial frequency of punishers is set to $z(0) = 0.5$. The  parameters  are  $N=50$, $K_m=10$, $\alpha = 0.05$, $\gamma = 1.5$, $\beta=3$, $r=2$ and $c=1$. 
 }  
\label{fig:8}  
\end{figure}

To quantify the advantage of the cooperators for finite  populations,  we consider the parameter setting of the simplex plot in the left panel of Fig.\ \ref{fig:5},  where  any  orbit starting from an initial condition inside the simplex leads to the all-defectors fixed point.  The probabilities of fixation $\Pi_C$, $\Pi_D$, and $\Pi_P$  are approximated by the fractions of the runs that lead to the fixation of cooperators, defectors and punishers, respectively. In Fig.\  \ref{fig:8} we fix $z(0)=0.5$ and vary $x(0)$ in the interval $(0,0.5)$ for several population sizes. The total number of independent runs is $10^4$.    Since we never observed the fixation of the punishers (i.e., $\Pi_P = 0$), it is sufficient to show $\Pi_C$ in this figure.  The results show that fixation of cooperators is very likely for small populations.  Even for large populations, the cooperators are very likely to end up as winners, provided, somewhat counterintuitively, that their initial frequency is sufficiently low. This is a consequence of the peculiarity of the trajectories shown in the simplex plot in the left panel of Fig.\ \ref{fig:5}. As $M$ increases, the range of initial conditions $x(0)$  for which $\Pi_C$ is  not zero  shrinks to zero, as expected from  the analysis of the replicator equations.  More specifically, 
$x(0)$ decreases with $M^{-0.15}$, as shown in the right panel of Fig.\ \ref{fig:8}. This result implies that, for example, for $x(0) = 0.2 M^{-0.15}$ and $z(0)=0.5$, the cooperators are virtually certain to fixate regardless of the size of the population.  

\begin{figure}[th] 
\centering
 \includegraphics[width=1\columnwidth]{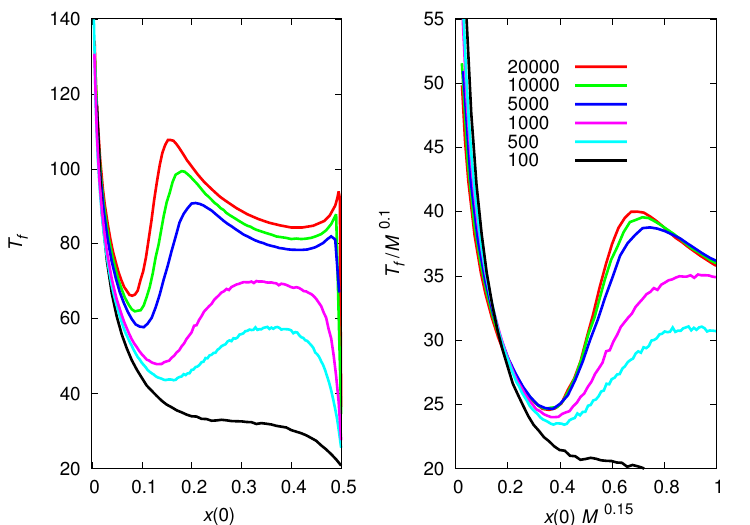}  
\caption{Mean time for fixation of cooperators or defectors $T_f$ as a function of the initial frequency of cooperators  (left panel) for populations of sizes  $M=100, 500, 1000, 5000, 10000$ and $20000$. The right panel shows the suitability of the scaling assumption (\ref{Tscal}). The color convention is the same for both panels. The initial frequency of punishers  is set to $z(0) = 0.5$. The  parameters  are  $N=50$, $K_m=10$, $\alpha = 0.05$, $\gamma = 1.5$, $\beta=3$, $r=2$ and $c=1$.  $T_f$ is dimensionless and  is measured in Monte Carlo steps.
 }  
\label{fig:9}  
\end{figure}

Another important quantity to characterize the stochastic dynamics is the mean  fixation time $T_f$, i.e., the mean time for the dynamics to reach the all-cooperators or the all-defectors   absorbing  configurations, which is shown in Fig.\ \ref{fig:9}.  We recall that $T_f$ is dimensionless and  is measured in Monte Carlo steps.  For small $x(0)$ and large $M$, the cooperators fixate in practically all runs (see Fig.\ \ref{fig:8}) and  the fixation time decreases  with increasing $x(0)$, but undergoes an abrupt increase when it is the turn of the defectors to fixate with certainty.  In this regime, $T_f$ decreases slowly as $x(0)$ increases, which implies that $y(0)$ decreases.   In the region of interest, i.e., for $\Pi_C \approx 1$ and large $M$, the scaling assumption
\begin{equation}\label{Tscal}
T_f = M^b f \left  [ x(0)M^a  \right ]
\end{equation}
with $a =0.15$ and  $b=0.1$  fits the data  well, as shown in the right panel of   Fig.\ \ref{fig:9}. Here  $f(u)$ is a  scaling function that satisfies $\lim_{u \to 0} f(u) \to \infty$. Note that the scaling form is obeyed only for very large $M$ due to the low values of the exponents $a$ and $b$.

\begin{figure}[th] 
\centering
 \includegraphics[width=1\columnwidth]{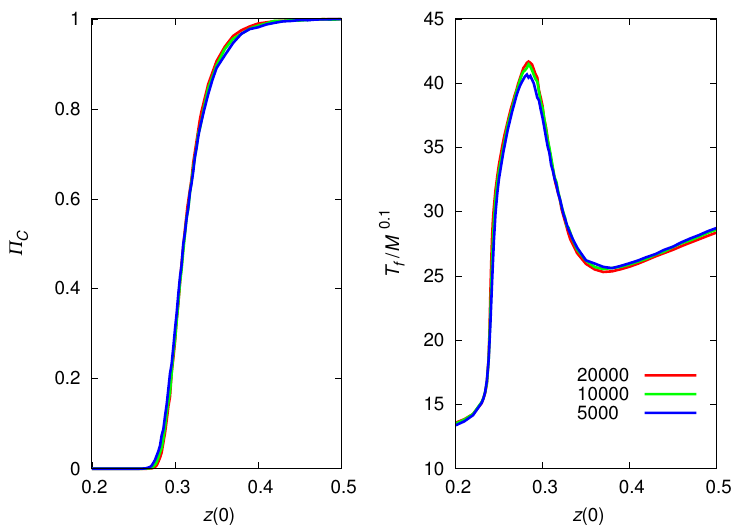}  
\caption{Probability of fixation of cooperators $\Pi_C$  (left panel)  and  scaled mean time for fixation $T_f/ M^{0.1}$ (right panel)  as a function of  initial frequency of punishers $z(0)$ for $M= 5000, 10000$ and $20000$.  The color convention is the same for both panels.
The initial frequency of cooperators is set to $x(0) = 0.2M^{-0.15}$. The  parameters  are  $N=50$, $K_m=10$, $\alpha = 0.05$, $\gamma = 1.5$, $\beta=3$, $r=2$ and $c=1$.  $T_f$ is dimensionless and  is measured in Monte Carlo steps.
 }  
\label{fig:10}  
\end{figure}

 To understand  how the fixation probability of the cooperators is affected by  the choice of the initial frequency of punishers, in Fig.\ \ref{fig:10}  we set $x(0) = 0.2 M^{-0.15}$ and vary $z(0)$ for different population sizes $M$. As shown before, this scaling guarantees  the fixation of the cooperators   for  $z(0)=0.5$ in the limit of large  $M$. The results indicate that there is a threshold for $z(0)$  below which the cooperators have no chance against the defectors. At this threshold,  the mean fixation time is maximal.

The finite population analysis gives unexpected results only for the scenario shown in the left panel of Fig.\ \ref{fig:5}, where the replicator equations predict that the cooperators win over the punishers. In the bistability scenario of the right panel of Fig.\ \ref{fig:4}, the initial conditions must be set very close to the separatrix in order for the results of the stochastic dynamics to differ from the deterministic predictions, as shown in Fig.\ \ref{fig:7}. As expected, in the scenario where the defectors win over the punishers in the deterministic regime, shown in the right panel of Fig.\ \ref{fig:5}, the result of the stochastic dynamics is the fixation of the defectors.

%
 \section{Discussion}\label{sec:conc}
 %

In contrast to most public goods game studies, which, following the seminal experimental studies in behavioral economics \cite{Fehr_2000,Fehr_2002}, consider small groups of players, we focus here on large groups (i.e., communities), since the setting up and maintenance of a sanctioning institution requires a large number of committed individuals.  Peer punishment (i.e., vigilantism) is unlikely in communities not only because of the high individual cost of exposure to retaliation, but also because of a well-known phenomenon in social psychology known as the diffusion of responsibility effect: people often fail to cooperate or punish offenders because they believe that others will or should take responsibility for doing so \cite{Darley_1968}.

In the community context, we note the  striking similarity between the replicator equation approach to public goods games and Wilson's model of  natural selection of populations and communities  \cite{Wilson_1975}, which focuses on individual traits that affect  the fitness  of other members of the population, such as behaviors that alter the environment (e.g., pollution and resource depletion).  More pointedly, in Wilson's  model  the fitness of individuals is determined locally, taking into account interactions within their trait groups, but their chances of reproduction are dictated by competition in the population at large. This is virtually identical to the imitation dynamics of the $N$-person games: a player's payoff is determined by the composition of the group of $N$ players to which she belongs, but the decision to change her strategy or not is determined by comparison with the payoff of another player randomly selected from the population, as described in Section \ref{sec:model}.  Another point of similarity between the more recent interdisciplinary game theoretic approach to social dilemmas and the traditional biological approach to the evolution of altruism that is worth noting is that the payoffs of cooperators and defectors in the others-only scenario in the absence of punishment (i. e, setting $\beta=0$ in Eqs. (\ref{fC}) and (\ref{fD})) are identical to the fitness used in Hamilton's classic work on the innate social behavior of humans \cite{Hamilton_1975}, a point overlooked in the recent literature \cite{Hannelore_2010}.

The study of large groups in finite as well as infinite populations is done in a variant of the standard formulation of the $N$-person prisoner's dilemma with pool punishment, where there is no overpunishment, i.e, the punishments for the different types of offenses are fixed, and a minimum number $K_m=\zeta N$ of punishers is required to form the sanctioning institution, which must be maintained regardless of the presence or absence of offenders. This constant maintenance cost is actually what distinguishes institutional punishment from peer punishment. The burden on punishers is reduced by assuming that the cost of maintenance is shared among punishers once their number exceeds $K_m$. All of these features have already been considered in the literature in a variety of different contexts (see, e.g., \cite{Dercole_2013,Vasconcelos_2013,Gois_2019,Boyd_2010,Pacheco_2009}), but here we have integrated  them to offer a more complete model of  institutional punishment.

As expected, the replicator equations which describe the imitation dynamics in the infinite population limit (i.e., $M \to \infty$) predict that the all-defectors solution is always stable and that the all-cooperators solution is always unstable.  The all-punishers solution is  stable  if the per capita cost to punishers of contributing to the sanctioning institution is less than the penalty to cooperators, and this cost plus the cost of contributing to public goods is less than the penalty to defectors.   In this case, there is a lower threshold $z^*$ for the frequency of punishers in the population at large that guarantees the maintenance of the sanctioning institution and, consequently, the disappearance of both types of offenders, i.e., cooperators and defectors. This threshold has a non-trivial dependence on the community size $N$ (see Figs.\ \ref{fig:1} and \ref{fig:2}): $z^*$ increases with increasing $N$ when the per capita contribution to the sanctioning institution $\gamma$ is small, and decreases when it is large. 
Sharing the cost of maintaining the sanctioning institution among the punishers once their number exceeds $K_m$ has some notable quantitative effects. To appreciate this, note that the scenario where there is no cost sharing and the punishers' contribution to the sanctioning institution is $\gamma$, regardless of their number, is obtained by setting $\zeta=1$ in Eqs.\ (\ref{lpu1}) and (\ref{lpu2}), which determine the eigenvalues of the community matrix for the all-punishers solution. Thus, sharing costs (i.e.,  $\zeta < 1$) increases the region of stability of the all-punishers fixed point in the parameter space. Similarly,  Eqs.\ (\ref{znocinf}) and (\ref{znodinf})   show that cost sharing  also increases the  size of the  domain of attraction of the all-punishers fixed point in the case of bistability.

We note that the scenario of bistability between punishers and defectors, i.e., the dependence of the equilibrium solution on the initial frequencies of the strategies, is a typical result of well-mixed populations, which is usually not observed in spatial public goods games, where players are fixed at the sites of a lattice and interactions are only possible between neighbors (see, e.g., \cite{Helbing_2010a,Helbing_2010b,Wang_2024,Szolnoki_2011a}).  In fact, spatial evolutionary games can be quite different from their well-mixed (mean-field) counterparts.
For example, in the spatial  2-person prisoner's dilemma, by randomizing the location of the players, there is a high probability that two (or more) cooperators will be neighbors, resulting in changing spatial patterns where both cooperators and defectors persist indefinitely \cite{Nowak_1992}, whereas the well-mixed population approach predicts the rapid extinction of cooperators.  Also, the random assignment of mutually dependent players in the same spatial neighborhood allows full invasion of resident populations when mean-field theory predicts bistability \cite{Rosas_2002}. However, in the institutional punishment scenario considered in our paper, where large group sizes $N$ and a minimum number of punishers $K_m = \zeta N$ are required to form the punishing institution, spatial localization is unlikely to solve the problem of  the dependence on initial frequencies  leading to bistability in a well-mixed population: we need to have at least  $K_m$  punishers  in the initial population setting, and the probability that they will all be assigned  to sites in the same extended neighborhood is vanishingly small.

The imitation stochastic dynamics \cite{Fontanari_2024b,Traulsen_2005,Sandholm_2010} for finite populations produces some unexpected results in the case where the replicator equations predict that cooperators win over punishers, who in turn win over defectors if their initial frequency is sufficiently high (see left panel of Fig.\ \ref{fig:5}). We find that demographic noise helps to fixate cooperators even for infinitely large populations, provided, somewhat counterintuitively, that the initial fraction of cooperators vanishes with some small power of the population size.  Of course, since a population of cooperators can be invaded by a single defector regardless of its size, this is not a robust scenario for maintaining cooperation.  

Our approach follows the Hobbesian view that, human beings being the selfish creatures that they are, life would be "solitary, poor, nasty, brutish, and short", were it not for the existence of repressive mechanisms \cite{Hobbes_1651}. From a game theory perspective, the main difficulty in establishing this scenario, which seems to be the current state of affairs, is the need for a minimum number of individuals willing to contribute to the establishment of a sanctioning institution. We mention, however, that there are alternative scenarios that actually seem to describe early human societies more accurately, where cooperation is the norm and sanctioning institutions are essentially nonexistent \cite{Graeber_2021}.

\section*{Acknowledgments}
JFF is partially supported by  Conselho Nacional de Desenvolvimento Ci\-en\-t\'{\i}\-fi\-co e Tecnol\'ogico  grant number 305620/2021-5.

\section*{Declaration of interest:} None.

\appendix

\renewcommand{\theequation}{A.\arabic{equation}}
\setcounter{equation}{0}
\setcounter{figure}{0}

\section{}\label{ref:A}

Here we show how the  sums in Eqs. (\ref{piC}), (\ref{piD}), and (\ref{piP}) can be explicitly evaluated in the limit of large group size $N$,  and hence large  $K_m =  \zeta N$,  to derive  the  discontinuous  expected payoffs given  in Eqs. (\ref{piCinf}), (\ref{piDinf}),  and (\ref{piPinf}).

The difficulty to evaluate   the expected payoff of a cooperator $\pi^C$ given in Eq. (\ref{piC}) is to calculate the sum
\begin{equation}
\Sigma_C = \sum_{K=0}^{K_m-1}\binom{N-1}{K}  z^K  (1-z)^{N-1-K}  
\end{equation}
where the summand is the binomial distribution with parameters $N-1$ and $z$. For large $N$ and $z$ not too small we can replace this distribution by a Gaussian distribution with mean $\mu =  N z$ and variance $\sigma^2 = Nz(1-z)$,
\begin{eqnarray}
\Sigma_C &  \approx  & \int_0^{K_m} \frac{dk}{\sqrt{2 \pi \sigma^2}} \exp \left [- \frac{(k - \mu)^2}{2\sigma^2} \right ]  \nonumber \\
 &  \approx  & \frac{1}{\sqrt{\pi}} \int_{\frac{-\mu}{\sqrt{2 \sigma^2}}}^{\frac{K_m-\mu}{\sqrt{2 \sigma^2}}}  dy  \exp \left [- y^2  \right ] \nonumber \\
  &  \approx  & \frac{1}{2} \mbox{erf} \left [ \frac{K_m - \mu}{\sqrt{2\sigma^2}}  \right ]
+
\frac{1}{2} \mbox{erf} \left [ \frac{ \mu }{\sqrt{2\sigma^2}}  \right ] .
\end{eqnarray}
Plugging the explicit expressions for $K_m$, $\mu$, and $\sigma^2$ into this equation yields
\begin{equation}\label{SC1}
\Sigma_C \approx   \frac{1}{2}  \mbox{erf} \left [  \frac{ \sqrt{N}(\zeta -z)}{\sqrt{2z(1-z)}}   \right ]  +   \frac{1}{2}
\mbox{erf} \left [  \frac{ \sqrt{N} z }{\sqrt{2z(1-z)}}   \right ] ,
\end{equation}
so now we can easily take the limit $N \to \infty$.  Noticing that $\lim_{x \to \infty} \mbox{erf}(x) \to 1$ and $\lim_{x \to -\infty} \mbox{erf}(x) \to -1$, we get
\begin{equation}\label{SC2}
\Sigma_C=
    \begin{cases}
     1 & \text{if $z < \zeta $ }\\
     	1/2 & \text{if $z =\zeta $ } \\
     0 & \text{f $z > \zeta $}.
    \end{cases}       
\end{equation}
Inserting this result into Eq.\ (\ref{piC}) yields the cooperators' mean payoff in the limit of infinite group sizes, Eq.\ (\ref{piCinf}). Since $\Sigma_C$ also appears in Eq.\ (\ref{piD}), the mean payoff of the defectors in this limit, given in Eq.\ (\ref{piDinf}), is also immediately obtained. Finally, the two sums appearing in Eq.\ (\ref{piP}) are identical to $\Sigma_C$ in the limit $N \to \infty$, since $N-1 \approx  N$ and $K_m-2 \approx  K_m$ in this limit, leading directly to the mean payoff of the punishers given in Eq.\ (\ref{piPinf}).

Note that the variable $z$ is determined by the replicator equations (\ref{x}) and (\ref{z}), and so we have no a priori control over the values it takes in the interval $[0,1]$. In particular, by solving the equations  $\pi^C(z^*) = \pi^D(z^*) = \pi^P(z^*)$  for finite $N$ (i.e.,  numerically evaluating  $\Sigma_C$ and the other sums appearing in the equations for the strategies' mean payoffs) we find that  the equilibrium coexistence solution $z^*$  tends to $\zeta$ as $N$ increases. More precisely, we find that $z^* - \zeta \sim 1/\sqrt{N}$. This implies that the argument of the error function  in the first term of the rhs of Eq. (\ref{SC1})  does not diverge in the limit $N \to \infty$. We deal with this problem in the analysis of the coexistence solution by introducing a new variable  $\eta = \sqrt{N}( \zeta -z^*)$  and deriving an equation for it  (see Eq. (\ref{etac})).

\end{document}